\documentclass[amssymb, amsmath,nobibnotes, aps, prl]{revtex4}

\usepackage{stmaryrd}
\usepackage{amssymb}
\usepackage{amsmath}
\usepackage{epsfig}
\begin{document}
\large

\title{Relaxation of Pseudo pure states : The Role of Cross-Correlations}
\author{Arindam Ghosh}
\author{Anil Kumar}
\altaffiliation{DAE-BRNS Senior Scientist.}
\email{anilnmr@physics.iisc.ernet.in}
\affiliation{NMR Quantum Computation and Quantum Information Group\\ Department of Physics and Sophisticated Instruments
Facility, Indian Institute of Science,
Bangalore-560012, India}

\begin{abstract}
In Quantum Information Processing by NMR one of the major challenges is relaxation or decoherence. Often it is found 
that the equilibrium mixed state of a spin system is not suitable as an initial state for computation and a definite initial 
state is required to be prepared prior to the computation. 
As these preferred initial states are non-equilibrium states, they are not stationary and are destroyed with time as the spin 
system relaxes toward 
its equilibrium, introducing error in computation. Since it is not possible to cut off the relaxation processes completely, 
attempts are going on to develop alternate 
strategies like Quantum Error Correction Codes or Noiseless Subsystems. Here we study 
the relaxation behavior of various Pseudo Pure States and analyze the role of Cross terms between different relaxation 
processes, known as Cross-correlation. It is found that while cross-correlations accelerate 
the relaxation of certain pseudo pure states, they retard that of others.\\

\end{abstract}
\maketitle

\section{I. Introduction}

Quantum information processing (QIP) often requires pure state as the initial state \cite{preskill,chuangbook}. Shor's prime 
factorizing algorithm \cite{shor}, Grover search algorithm \cite{grover} are few examples. Creation of pure state in NMR is 
not easy due to small gaps between nuclear magnetic energy levels and demands unrealistic experimental conditions like near absolute 
zero temperature or extremely high magnetic field. This problem has been circumvented by creating a Pseudo Pure State (PPS).
While in a pure state all energy levels except one have zero populations, in a PPS all levels except one have equal
populations. Since the uniform background populations do not
contribute to the NMR signal, such a state then mimics a pure state. Several methods of 
creating PPS have been developed like spatial averaging 
\cite{cory97,cory98}, logical labeling \cite{chuang97,chuang98,kavita00,kavita1}, temporal averaging \cite{chuangtemp}, spatially 
averaged logical labeling 
technique (SALLT) \cite{maheshpra01}. However pseudo pure state, as well as pure states are not stationary and are destroyed with 
time as the spin system relaxes toward equilibrium. 
In QIP there are also cases where
one or more qubits are initialized to a suitable state at the beginning of the computation and are used as storage or memory qubits 
at the end of the computation performed on some other qubits\cite{chuangdelay}. In these cases it is important for memory qubits to be in the 
initialized state till the time 
they are in use since deviation from the initial state adds error to the output result. Since it is not possible to stop 
decay of
a state which is away from equilibrium, alternate strategies like Quantum Error Correction \cite{qec}, Noiseless
subspace \cite{ns1,ns2} are being tried. Recently Sarthour et al.\cite{quadrel} has reported a detailed study of relaxation of pseudo
pure states and few other states in a quadrupolar system. Here we experimentally examine the lifetime of various pseudo pure states in 
a weakly J-coupled two qubit system. We find that cross
terms (known as cross-correlation) between different pathways of relaxation of a spin can retard the relaxation of certain
PPS and accelerate that of others.\\\\
In 1946 Bloch formulated the behavior of populations or longitudinal magnetizations when they are perturbed from the 
equilibrium \cite{bloch}. 
The recovery toward equilibrium is exponential for a two level system and for a complex system the recovery involves 
several time constants \cite{redfield}. For complex systems the von Neumann-Liouville equation \cite{von,abragam}
describes mathematically the time evolution of the density matrix in the magnetic resonance phenomena. 
For system having more than one spin
the relaxation is described by a matrix called the relaxation matrix whose elements are linear combinations of spectral densities, which 
in turn are Fourier transforms of time correlation function \cite{anilreview} of the fluctuations of the various interactions responsible
for relaxation. There exist several different mechanisms for relaxation, such as, 
time dependent dipole-dipole(DD) interaction, chemical shift anisotropy(CSA), quadrupolar interaction and spin rotation interaction \cite{anilreview}. 
The correlation
function gives the time correlations between different values of the interactions. The final correlation function has two major parts, 
namely the `Auto-correlation' 
part which gives at two different times the correlation between the same relaxation interaction and the 
`Cross-correlation' part  which gives the time correlation between two different relaxation interactions. The mathematics of cross
correlation can be found in detail, in works of Schneider \cite{sch1,sch2}, Blicharski \cite{blicharski} and Hubbard \cite{hubbard}.
Recently a few models have been suggested to study the decoherence of the quantum coherence, the off-diagonal elements in density matrix
\cite{zurek,cory03}. It can be shown that in absence of r.f. pulses and under secular approximation 
the relaxation of
the diagonal and the off-diagonal elements of the density matrix  are independent \cite{redfield}. Here we study the
longitudinal relaxation that is the 
relaxation of the diagonal elements of the density matrix and the role of cross-correlations in it.
\newpage
\section{II. Theory}
\subsection{A. The Pseudo Pure State (PPS) : In terms of magnetization modes}
\begin{center}
\begin{figure}
\vspace*{-1cm}
\epsfig{file=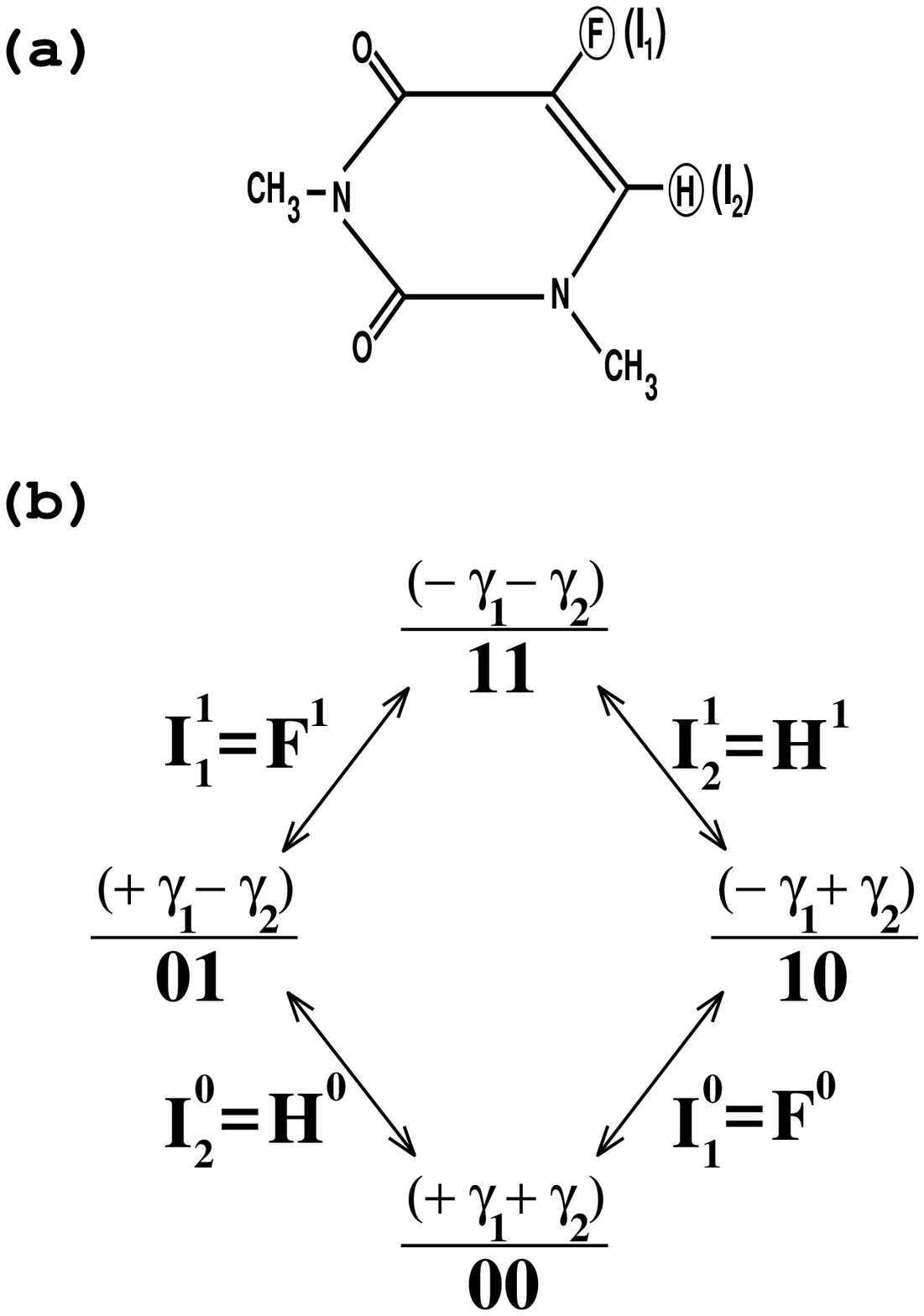,angle=0,width=10cm}
\caption{}
\label{eqlev}
\end{figure}
\end{center}

\begin{center}
\begin{figure}
\vspace*{-1cm}
\epsfig{file=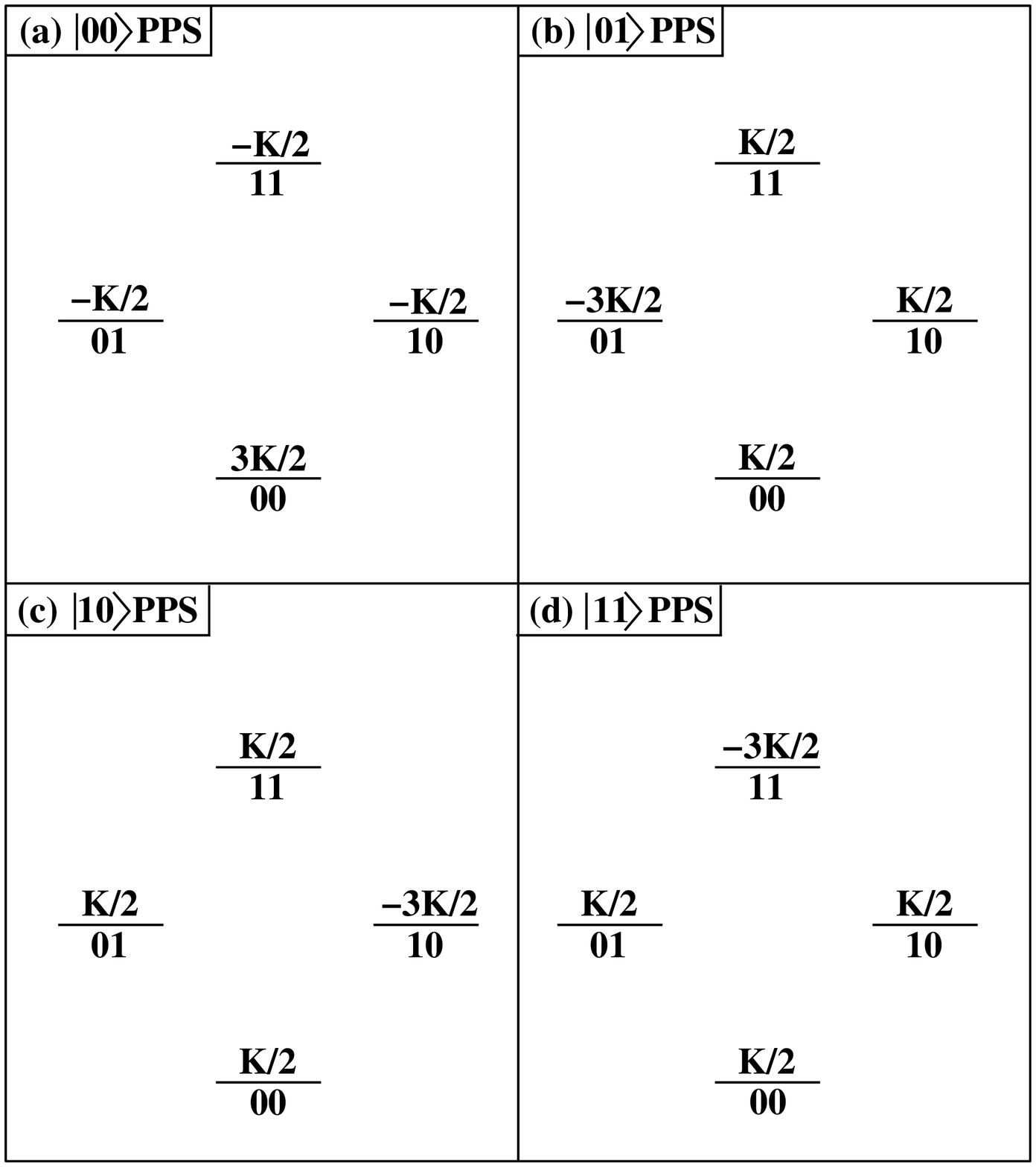,angle=0,width=16cm}
\caption{}
\label{ppslev}
\end{figure}
\end{center}

In terms of magnetization modes the equilibrium density matrix of a two spin system 
is given by \cite{cory98,abragam,ernstbook,jonesgate}[Fig.\ref{eqlev}],
\begin{eqnarray}
\chi_{eq} = \gamma_1 I_{1Z} + \gamma_2 I_{2Z} 
\label{eqd}
\end{eqnarray}
where $\gamma_1$ and $\gamma_2$ are gyro-magnetic ratios of the two spins $I_1$ and $I_2$ respectively. 
The density matrix of a general state can be written as,
\begin{eqnarray}
\chi = \pm \alpha \gamma_1 I_{1Z} \pm \beta \gamma_2 I_{2Z} \pm \nu \gamma_1 2I_{1Z}I_{2Z}]
\label{general}
\end{eqnarray}
which for the condition $\alpha \gamma_1$=$\beta \gamma_2$=$\nu \gamma_1$=K, corresponds to the density matrix of a PPS
given by \cite{cory98},
\begin{eqnarray}
\chi_{pps} = K[\pm I_{1Z} \pm I_{2Z} \pm 2I_{1Z}I_{2Z}]
\label{pps}
\end{eqnarray}
where, K is a constant, the value of which depends on the method of creation of PPS. \\
The first two terms in the right hand side in Eq.\ref{general} and Eq.\ref{pps} are the single spin order modes for the first and second spin respectively while the last term is 
the two spin order mode of the two spins \cite{cory98}. Choosing properly the signs of the modes, the various PPS of a
two-qubit system are,
\begin{eqnarray}
\chi_{pps}^{00} =  K[+ I_{1Z} + I_{2Z} + 2I_{1Z}I_{2Z}]\nonumber \\   \chi_{pps}^{01} =  K[- I_{1Z} + I_{2Z} +2I_{1Z}I_{2Z}] \nonumber \\
\chi_{pps}^{10} =  K[+ I_{1Z} - I_{2Z} + 2I_{1Z}I_{2Z}]\nonumber \\   \chi_{pps}^{11} =  K[+ I_{1Z} + I_{2Z} -2I_{1Z}I_{2Z}]
\end{eqnarray}
The relative populations of the states for different PPS are shown in Fig. \ref{ppslev}. As seen in Eq.2, in PPS the coefficients of the all three modes are equal. 
On the other hand equilibrium density matrix
does not contain any two spin order mode. To reach Eq.\ref{pps} starting from Eq.\ref{eqd}, the two spin order mode has to be created and
 at the same time the coefficients of all the modes have to be made equal.

\subsection{B. Relaxation of Magnetization Modes}
The equation of motion of modes M is given by \cite{anilreview},
\begin{eqnarray}
- \frac{d}{dt} \vec{M}(t) = \hat{\Gamma} [\vec{M}(t) - \vec{M}(\infty)]
\label{magmode}
\end{eqnarray}
where $\hat{\Gamma}$ is the relaxation matrix and $\vec{M}(\infty)$ is the equilibrium values of a mode.
For a weakly coupled two-spin system relaxing via mutual dipolar interaction and the CSA relaxation, the two dominant mechanism of
relaxation of spin half nuclei in liquid state, the above equation takes the form,
\begin{eqnarray}
- \frac{d}{dt} 
\left [
\begin{array}{c}
I_{1Z}(t)\\
I_{2Z}(t)\\
2I_{1Z}I_{2Z}(t)
\end{array} 
\right ]
= 
\left [
\begin{array}{ccc}
\rho_1 & \sigma_{12} & \delta_{1,12}\\
\sigma_{12} & \rho_2 & \delta_{2,12}\\
\delta_{1,12} & \delta_{2,12} & \rho_{12}
\end{array}
\right ]
\cdot
\left [
\begin{array}{c}
I_{1Z}(0)-I_{1Z}(\infty)\\
I_{2Z}(0)-I_{2Z}(\infty)\\
2I_{1Z}I_{2Z}
\end{array}
\right ]
\label{relax}
\end{eqnarray}
where $\rho_i$ is the self relaxation rate of the single spin order mode of spin $\bf i$, $\rho_{ij}$ is the self
relaxation rate of the two spin order mode of spin $\bf i$ and $\bf j$, $\sigma_{ij}$ is the cross-relaxation (Nuclear Overhouser Effect,
NOE) rate between spins $\bf i$ and $\bf j$ and 
$\delta_{i,ij}$ 
is the cross-correlation term between CSA relaxation of spin $\bf i$ and the dipolar relaxation between the spins $\bf i$ and $\bf j$. 
$\rho$ and $\sigma$ involve only the 
auto-correlation terms 
and $\delta$ involves only the cross-correlation terms\cite{anilreview}. Magnetization 
modes of one order relaxes to other orders through
cross-correlation and in absence of it the relaxation matrix becomes block diagonal within each order. The relaxation of modes are 
in general dominated by their self relaxation $\rho$, but in case of samples having long $T_1$, the cross-correlation terms become 
comparable with self-relaxation and play an important role in relaxation of the spins.\\
The formal solution of Eq. \ref{magmode} is given by,
\begin{eqnarray}
\vec{M}(t) = \vec{M}(\infty) + [\vec{M}(0)-\vec{M}(\infty)] exp(-\hat{\Gamma}t)
\end{eqnarray}
As time evolution of various modes are coupled, a general solution of the above equation requires diagonalization of the relaxation
matrix. However, in the initial rate approximation Eq.7 can be written (for small values of t=$\tau$) as,
\begin{eqnarray}
\vec{M}(\tau)&=& \vec{M}(\infty) + [\vec{M}(0)-\vec{M}(\infty)][1-\hat{\Gamma}\tau] \cr
&=& \vec{M}(0) - \hat{\Gamma}\tau[\vec{M}(0)-\vec{M}(\infty)]
\label{initialapp}
\end{eqnarray}
This equation asserts that in the initial rate approximation (for low $\tau$), the decay or growth of a mode is
linear with time and the initial slope is proportional to the corresponding relaxation matrix element. If the modes are allowed to 
relax for a longer
time, their decay or growth deviates from the linear nature and adopts a multi-exponential behavior to finally reach the
equilibrium\cite{anilreview}.
\subsection{C. Relaxation of Pseudo pure state}
\begin{center}
\begin{figure}
\vspace*{-1cm}
\epsfig{file=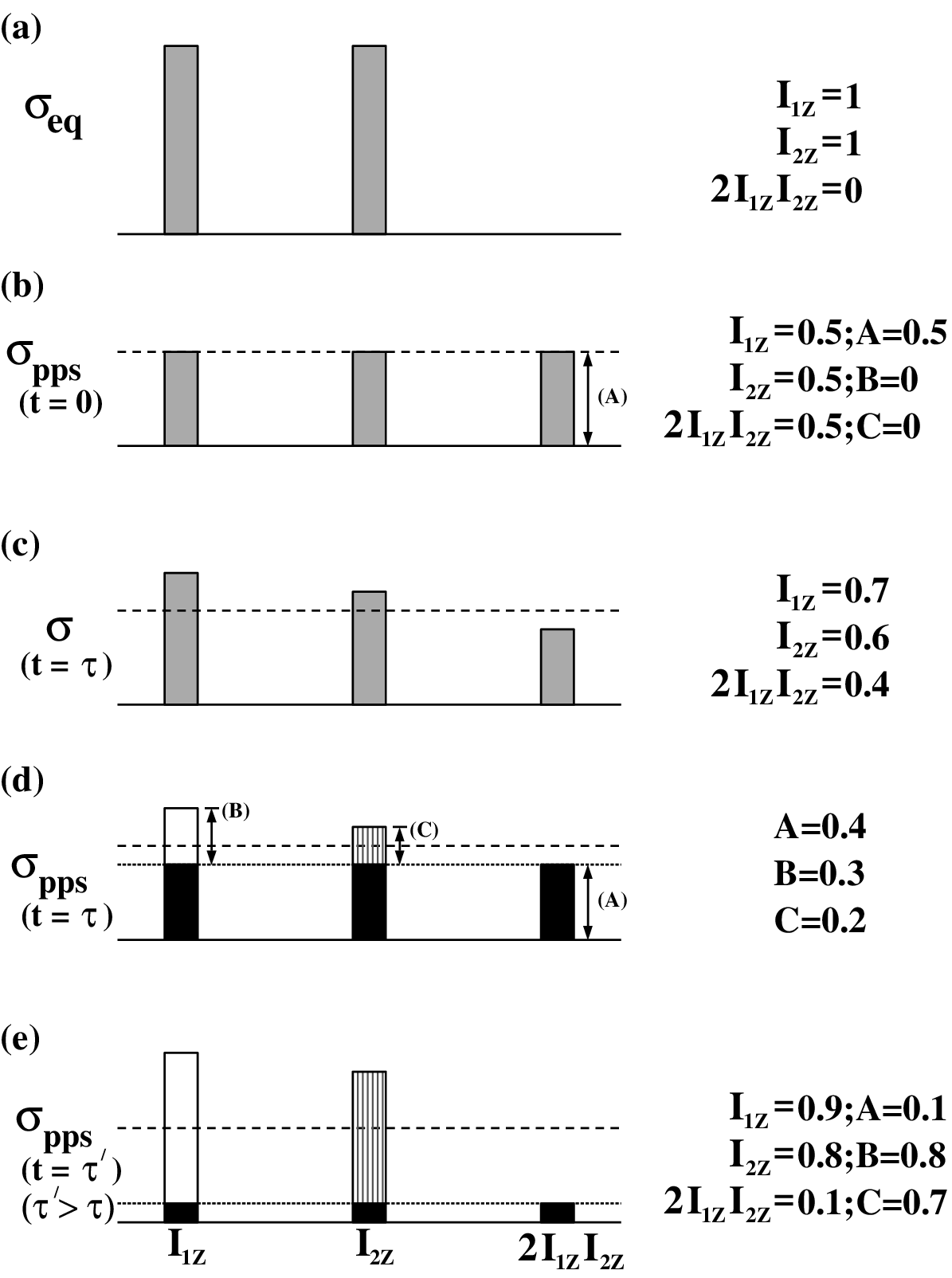,angle=0,width=16cm}
\caption{}
\label{ppsmode}
\end{figure}
\end{center}

Let a two qubit system be in $|00\rangle$ PPS at t=0. 
\begin{eqnarray}
\chi^{00}(0) = K[I_{1Z} + I_{2Z} + 2I_{1Z}I_{2Z}] 
\end{eqnarray}
After time t it will relax to,
\begin{eqnarray}
\chi^{00}(t)&=& (K + \Delta_1(t))I_{1Z} + (K + \Delta_2(t))I_{2Z} 
+ (K + \Delta_{12}(t))2I_{1Z}I_{2Z}
\label{e10}
\end{eqnarray}
where $\Delta_1(t)$,$\Delta_2(t)$ and $\Delta_{12}(t)$ are the time dependent deviations of respective modes from their
initial values. The deviation of the two spin order can be measured from spectrum of either spin. Eq.\ref{e10} can also be 
written as,
\begin{eqnarray}
\chi^{00}(t) = (K  + \Delta_{12}(t))[I_{1Z} + I_{2Z} + 2I_{1Z}I_{2Z}]
+ (\Delta_1(t) - \Delta_{12}(t))I_{1Z} 
+ (\Delta_2(t) - \Delta_{12}(t))I_{2Z} \label{chi001} 
\label{chi00}
\end{eqnarray}
The first term is the pseudo pure state with the coefficient decreasing in time while the other two terms are the
excesses of the single spin order modes with coefficients increasing in time. For other pseudo pure states Eq.\ref{chi001} 
becomes,
\begin{eqnarray}
\chi^{01}(t)&=&(K + \Delta_{12}(t))[-I_{1Z} + I_{2Z} + 2I_{1Z}I_{2Z}]
+(\Delta_1(t) + \Delta_{12}(t))I_{1Z}
+(\Delta_2(t) - \Delta_{12}(t))I_{2Z} \\
\chi^{10}(t) &=& (K + \Delta_{12}(t))[I_{1Z} - I_{2Z} + 2I_{1Z}I_{2Z}]
+(\Delta_1(t) - \Delta_{12}(t))I_{1Z}
+(\Delta_2(t) + \Delta_{12}(t))I_{2Z} \\
\chi^{11}(t) &=& (K - \Delta_{12}(t))[I_{1Z} + I_{2Z} - 2I_{1Z}I_{2Z}]
+(\Delta_1(t) + \Delta_{12}(t))I_{1Z}
+(\Delta_2(t) + \Delta_{12}(t))I_{2Z} 
\label{chi11}
\end{eqnarray}
In the initial rate approximation (using Eq.\ref{initialapp}) we obtain for the $|00\rangle$ pps,
\begin{eqnarray}
\Delta_1(\tau)& =& \tau[\rho_1(\gamma_1 -K) +\sigma_{12}(\gamma_2 -K) -K \delta_{1,12}] \label{d1}\\
\Delta_2(\tau)& =& \tau[\sigma_{12}(\gamma_1 -K) +\rho_{1}(\gamma_2 -K) -K \delta_{2,12}] \\
\Delta_{12}(\tau)& =& \tau[\delta_{1,12}(\gamma_1 -K) +\delta_{2,12}(\gamma_2 -K) -K \rho_{12}] \label{d2}
\end{eqnarray}
Let the coefficients of the PPS term and the two single spin order modes $I_{1Z}$ and $I_{2Z}$ in Eq.\ref{chi00} be called 
as $\mathcal{A}$,$\mathcal{B}$ and $\mathcal{C}$ respectively. Fig.\ref{ppsmode} schematically shows the time evolution of the 
coefficients $\mathcal{A}$,$\mathcal{B}$ and $\mathcal{C}$
for $|00\rangle$ PPS. Any coefficient for any PPS at any instant, is simply the initial value plus the total deviation due
to the auto and the cross-correlations. For example, ${\mathcal{A}}$ for $|00\rangle$ PPS at time $\tau$, 
is ${\mathcal{A}}^{00}(\tau)= K + {\mathcal{A}}^{00}_{auto}(\tau) +
{\mathcal{A}}^{00}_{cc}(\tau$), where K is the initial value, and ${\mathcal{A}}^{00}_{auto}(\tau)$
and ${\mathcal{A}}^{00}_{cc}(\tau)$ are the deviations at $\tau$ due to auto-correlation and cross-correlation
parts respectively. 
\subsubsection{(a) Contribution of auto-correlation terms to the deviation}

Putting 
the values of the deviations of different modes obtained from Eq.(\ref{d1}-\ref{d2}) in Eq.(\ref{chi001}-\ref{chi11}),
 we obtain the contribution 
only of auto-correlation terms to the deviation from initial value of the coefficients 
$\cal{A}$,$\mathcal{B}$ and 
$\mathcal{C}$ under initial rate approximation (at t=$\tau$) as,
\begin{eqnarray*}
{\mathcal{A}}_{auto}^{00}(\tau) = {\mathcal{A}}_{auto}^{01}(\tau) = {\mathcal{A}}_{auto}^{10}(\tau) = {\mathcal{A}}_{auto}^{11}(\tau) = 
-K \rho_{12} \tau
\nonumber \\
\end{eqnarray*}
\vspace{-2.0cm}
\normalsize
\begin{eqnarray}
{\mathcal{B}}_{auto}^{00}(\tau) = [\rho_{1}(\gamma_1 -K)+ \sigma_{12} (\gamma_2 -K) + K \rho_{12}]\tau& ;& 
{\mathcal{B}}_{auto}^{01}(\tau) = [\rho_{1}(\gamma_1 +K)+ \sigma_{12} (\gamma_2 -K) - K \rho_{12}]\tau \nonumber \\
{\mathcal{B}}_{auto}^{10}(\tau) = [\rho_{1}(\gamma_1 -K)+ \sigma_{12} (\gamma_2 +K) + K \rho_{12}]\tau& ;&
{\mathcal{B}}_{auto}^{11}(\tau) = [\rho_{1}(\gamma_1 -K)+ \sigma_{12} (\gamma_2 -K) + K \rho_{12}]\tau \nonumber \\
{\mathcal{C}}_{auto}^{00}(\tau) = [\sigma_{12}(\gamma_1 -K)+ \rho_2 (\gamma_2 -K) + K \rho_{12}]\tau& ;& 
{\mathcal{C}}_{auto}^{01}(\tau) = [\sigma_{12}(\gamma_1 +K)+ \rho_2 (\gamma_2 -K) + K \rho_{12}]\tau \nonumber \\
{\mathcal{C}}_{auto}^{10}(\tau) = [\sigma_{12}(\gamma_1 -K)+ \rho_2 (\gamma_2 +K) - K \rho_{12}]\tau& ;&
{\mathcal{C}}_{auto}^{11}(\tau) = [\sigma_{12}(\gamma_1 -K)+ \rho_2 (\gamma_2 -K) + K \rho_{12}]\tau
\end{eqnarray}
\large
It is evident that in absence of cross-correlations the $|00\rangle$ and $|11\rangle$ PPS relax at the same initial rate since 
${\mathcal{A}}_{auto}^{00}={\mathcal{A}}_{auto}^{11}$,
${\mathcal{B}}_{auto}^{00}={\mathcal{B}}_{auto}^{11}$ and ${\mathcal{C}}_{auto}^{00}={\mathcal{C}}_{auto}^{11}$. However
the same is not true for $|01\rangle$ and $|10\rangle$ PPS. 

\subsubsection{(b) Contribution of cross-correlation terms to the deviation}

The contribution by the cross-correlation terms is given by, 
\begin{eqnarray*}
{\mathcal{A}}^{00}_{cc}(\tau) = +[\delta_{1,12} (\gamma_1 -K)+ \delta_{2,12}(\gamma_2 -K)]\tau \nonumber \\ 
{\mathcal{A}}^{01}_{cc}(\tau) = +[\delta_{1,12} (\gamma_1 +K)+ \delta_{2,12}(\gamma_2 -K)]\tau \nonumber \\
{\mathcal{A}}^{10}_{cc}(\tau) = +[\delta_{1,12} (\gamma_1 -K)+ \delta_{2,12}(\gamma_2 +K)]\tau  \nonumber \\
{\mathcal{A}}^{11}_{cc}(\tau) = -[\delta_{1,12} (\gamma_1 -K)+ \delta_{2,12}(\gamma_2 -K)]\tau \nonumber \\ 
\end{eqnarray*}
\vspace*{-2.05cm}
\begin{eqnarray}
{\mathcal{B}}^{00}_{cc}(\tau) = - [\delta_{1,12} \gamma_1 + \delta_{2,12}(\gamma_2 -K)]\tau& ;& 
{\mathcal{B}}^{01}_{cc}(\tau) = + [\delta_{1,12} \gamma_1 + \delta_{2,12}(\gamma_2 -K)]\tau \nonumber \\
{\mathcal{B}}^{10}_{cc}(\tau) = - [\delta_{1,12} \gamma_1 + \delta_{2,12}(\gamma_2 +K)]\tau& ;&
{\mathcal{B}}^{11}_{cc}(\tau) = + [\delta_{1,12} \gamma_1 + \delta_{2,12}(\gamma_2 -K)]\tau\nonumber \\
{\mathcal{C}}^{00}_{cc}(\tau) = - [\delta_{1,12} (\gamma_1 -K) + \delta_{2,12} \gamma_2 ]\tau& ;& 
{\mathcal{C}}^{01}_{cc}(\tau) = - [\delta_{1,12} (\gamma_1 +K) + \delta_{2,12} \gamma_2 ]\tau\nonumber \\
{\mathcal{C}}^{10}_{cc}(\tau) = + [\delta_{1,12} (\gamma_1 -K) + \delta_{2,12} \gamma_2 ]\tau& ;&
{\mathcal{C}}^{11}_{cc}(\tau) = + [\delta_{1,12} (\gamma_1 -K) + \delta_{2,12} \gamma_2 ]\tau
\label{abc-cross} 
\end{eqnarray}  
The important
thing is that the presence of cross-correlation can lead to differential relaxation of all PPS. 
Positive cross-correlation rates $\delta_{1,12}$ and $\delta_{2,12}$, slow down the relaxation of all the 
three coefficients for 
$|00\rangle$ PPS since (${\mathcal{A}}^{00}(\tau)>
{\mathcal{A}}^{00}_{auto}(\tau),{\mathcal{B}}^{00}(\tau)<{\mathcal{B}}^{00}_{auto}(\tau),{\mathcal{C}}^{00}(\tau)<{\mathcal{C}}^{00}_{auto}(\tau)$),
while make the relaxation of
all three coefficients faster for
 $|11\rangle$ PPS since (${\mathcal{A}}^{11}(\tau)<
{\mathcal{A}}^{11}_{auto}(\tau),{\mathcal{B}}^{11}(\tau)>{\mathcal{B}}^{11}_{auto}(\tau),{\mathcal{C}}^{11}(\tau)>{\mathcal{C}}^{11}_{auto
}(\tau)$).
For $|01\rangle$ and $|10\rangle$ PPS cross-correlations give a mixed effect since
 (${\mathcal{A}}^{01}(\tau)>
{\mathcal{A}}^{01}_{auto}(\tau),{\mathcal{B}}^{01}(\tau)>{\mathcal{B}}^{01}_{auto}(\tau),{\mathcal{C}}^{01}(\tau)<{\mathcal{C}}^{01}_{auto}(\tau)$)
and (${\mathcal{A}}^{10}(\tau)>
{\mathcal{A}}^{00}_{auto}(\tau),{\mathcal{B}}^{10}(\tau)<{\mathcal{B}}^{10}_{auto}(\tau),{\mathcal{C}}^{10}(\tau)>{\mathcal{C}}^{10}_{auto}(\tau)$).
As the contributions of the
auto-correlation part for $|00\rangle$ and $|11\rangle$ PPS are equal, we have monitored the relaxation behavior only of $|00\rangle$ and
$|11\rangle$ PPS to study the effect of cross-correlations.\\\\

For samples having long $T_1$, where the cross-correlations becomes comparable with auto-correlation rates, the four PPS relax
with four different rates and the difference increases with the increased value of the cross-correlation terms.   
The three coefficients $\mathcal{A}$,$\mathcal{B}$ and $\mathcal{C}$ (normalized to the equilibrium line intensities) in terms of proton 
and fluorine line intensities for $|00\rangle$ PPS are,
\begin{eqnarray*}
{\mathcal{A}}(t) = \frac{H^0 (t) - H^1 (t)}{H^1 (\infty) + H^0 (\infty)} = \frac{F^0 (t) - F^1 (t)}{F^1 (\infty) + F^0 (\infty)} 
\end{eqnarray*}
\vspace*{-0.71cm}
\begin{eqnarray}
{\mathcal{B}}(t)& = &\frac{2H^1 (t)}{H^0 (\infty) + H^1 (\infty)} \nonumber \\
{\mathcal{C}}(t)& = &\frac{2F^1 (t)}{F^0 (\infty) + F^1 (\infty)}
\end{eqnarray}
and for $|11\rangle$ PPS are,
\begin{eqnarray*}
{\mathcal{A}}(t) = \frac{H^1 (t) - H^0 (t)}{H^1 (\infty) + H^0 (\infty)} = \frac{F^1 (t) - F^0 (t)}{F^1 (\infty) + F^0(\infty)}
\end{eqnarray*}
\vspace*{-0.71cm}
\begin{eqnarray}
{\mathcal{B}}(t)& = &\frac{2H^0 (t)}{H^0 (\infty) + H^1 (\infty)} \nonumber \\
{\mathcal{C}}(t)& = &\frac{2F^0 (t)}{F^0 (\infty) + F^1 (\infty)}
\end{eqnarray}
where, $H^0$ and $H^1$ are intensities of the two proton transitions, when the fluorine spin is respectively in state $|0\rangle$ and
$|1\rangle$. Similarly $F^0$ and $F^1$ are intensities of two fluorine transitions corresponding
to the proton spin being respectively in the state $|0\rangle$ and $|1\rangle$, as shown in Fig.\ref{eqlev} and Fig.\ref{allspec}. 
$H^0 (t)$ and $H^0 (\infty)$ give the $H^0$ line intensity respectively at time t and at equilibrium. Thus by monitoring the intensities 
of the two proton and two fluorine transitions as a function of time, one can calculate the coefficient $A(t)$ which is a measure of decay
of PPS.
\section{III. Simulation}
\begin{center}
\begin{figure}
\vspace*{-1cm}
\epsfig{file=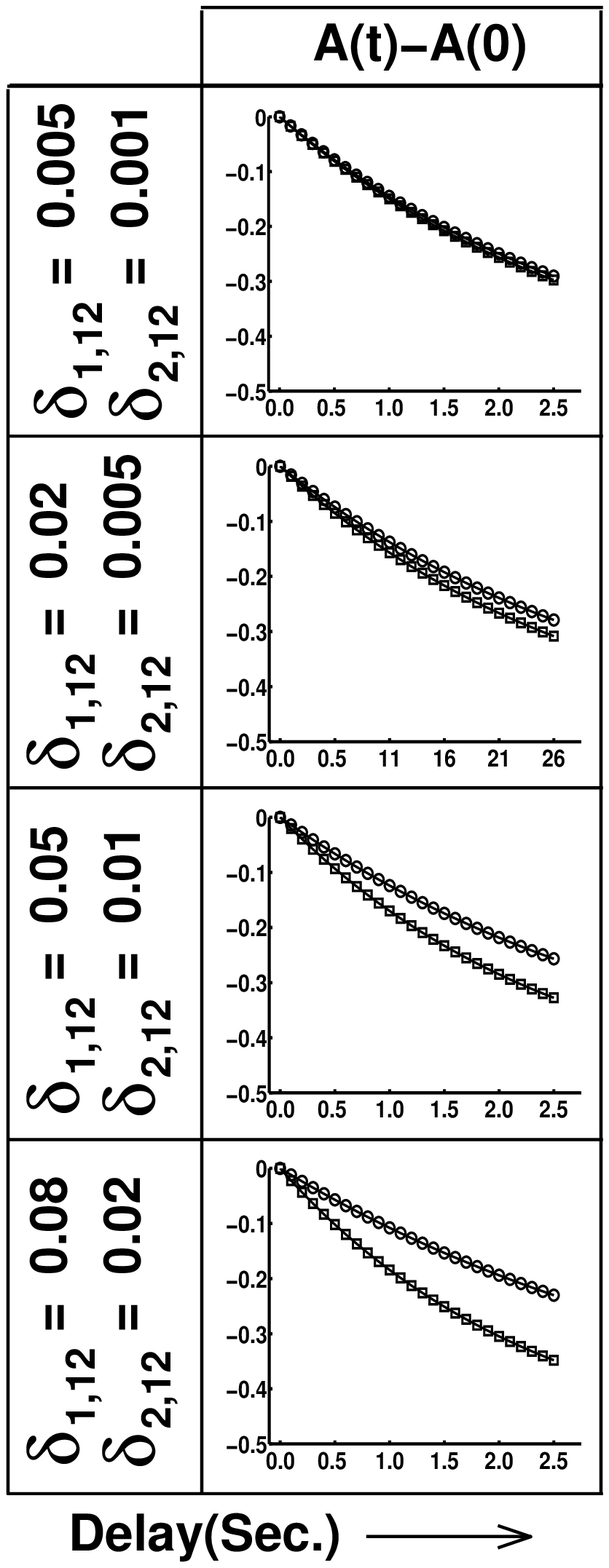,angle=0,height=23cm,width=9cm}
\caption{}
\label{Asimu}
\end{figure}
\end{center}

\begin{center}
\begin{figure}
\vspace*{-1cm}
\epsfig{file=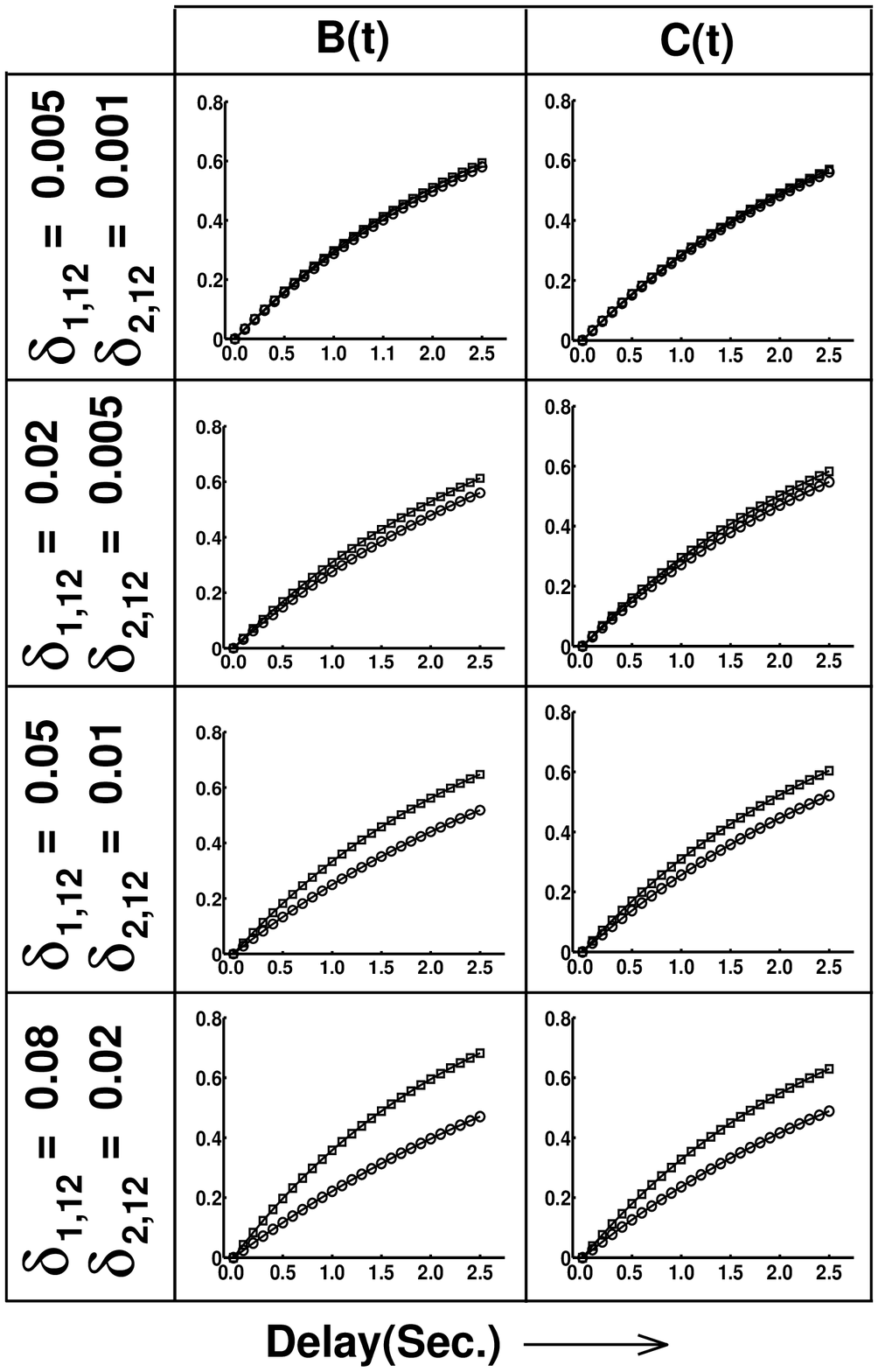,angle=0,height=23cm,width=12cm}
\caption{}
\label{BCsimu}
\end{figure}
\end{center}

\begin{center}
\begin{figure}
\vspace*{-1cm}
\epsfig{file=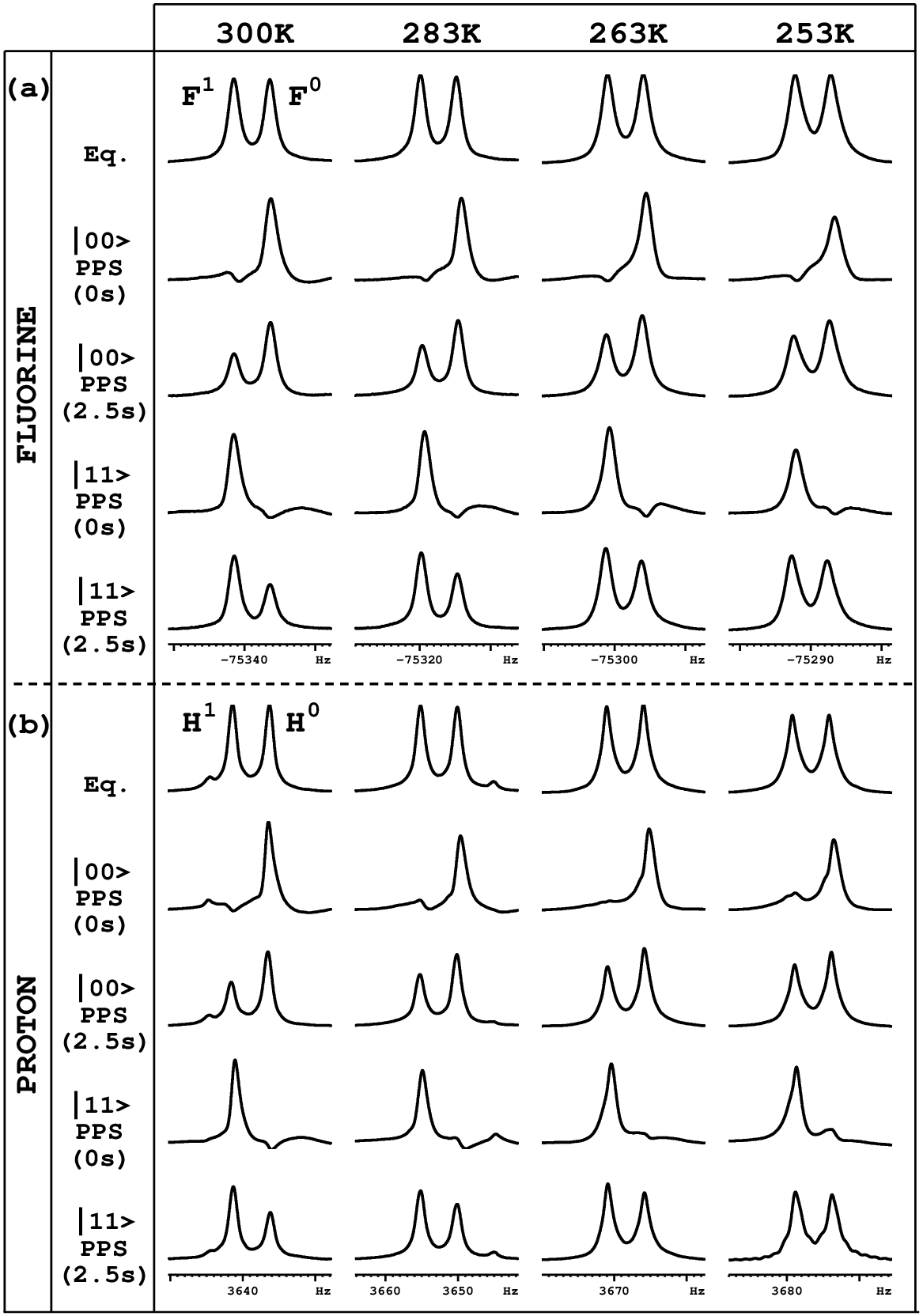,angle=0,width=16cm}
\caption{}
\label{allspec}
\end{figure}
\end{center}

Relaxation of the coefficients $\mathcal{A}$,$\mathcal{B}$ and $\mathcal{C}$ have been simulated using MATLAB, for a weakly coupled 
$^{19}F$-$^{1}H$ system. The relaxation matrix used for the simulation is,
\begin{eqnarray*}
\hat{\Gamma}
=
\left [
\begin{array}{ccc}
0.3125 & 0.02 & \delta_{1,12}\\
0.02 & 0.33 & \delta_{2,12}\\
\delta_{1,12} & \delta_{2,12} & 0.33
\end{array}
\right ]
\end{eqnarray*}
Fig.\ref{Asimu} shows the decay of coefficient $\mathcal{A}$ with time. ${\mathcal{A}}^{00}$ and ${\mathcal{A}}^{11}$ show no difference in
decay rate in absence of cross-correlation rates. As $\delta_{1,12}$ and $\delta_{2,12}$ are increased more and more difference in  decay
rate is observed. Fig.\ref{BCsimu} shows growth of coefficients $\mathcal{B}$ and $\mathcal{C}$. As $\delta_{2,12}$ is 
taken smaller than $\delta_{1,12}$, difference in decay rate between ${\mathcal{C}}^{00}$ and ${\mathcal{C}}^{11}$ is found to be less
than between ${\mathcal{B}}^{00}$ and ${\mathcal{B}}^{11}$.    
\section{IV. Experimental}
All the relaxation measurement were performed on a two qubit sample formed by one fluorine and one proton of 5-fluro 1,3-dimethyl uracil 
yielding an AX spin system with a J-coupling of 5.8 Hz.
Longitudinal relaxation time constants for $^{19}F$ and $^1H$ are 6 and 7.2 Sec respectively at room temperature (300K). All the experiments were performed in a Bruker DRX 500 MHz spectrometer where the resonance frequencies for
$^{19}F$ and $^1H$ are 470.59 MHz and 500.13 MHz respectively. The Pseudo-pure 
 state was prepared by spatial averaging method
using J-evolution \cite{cory98}.   
Relaxation of all the three
coefficients for $|00\rangle$ and $|11\rangle$ PPS has been calculated. Since auto-correlations contribute equally to the 
relaxation of 
these two PPS, any difference in relaxation rate can be attributed to cross-correlation rates. \\
 
\begin{center}
\begin{figure}
\vspace*{1cm}
\hspace*{-1cm}
\epsfig{file=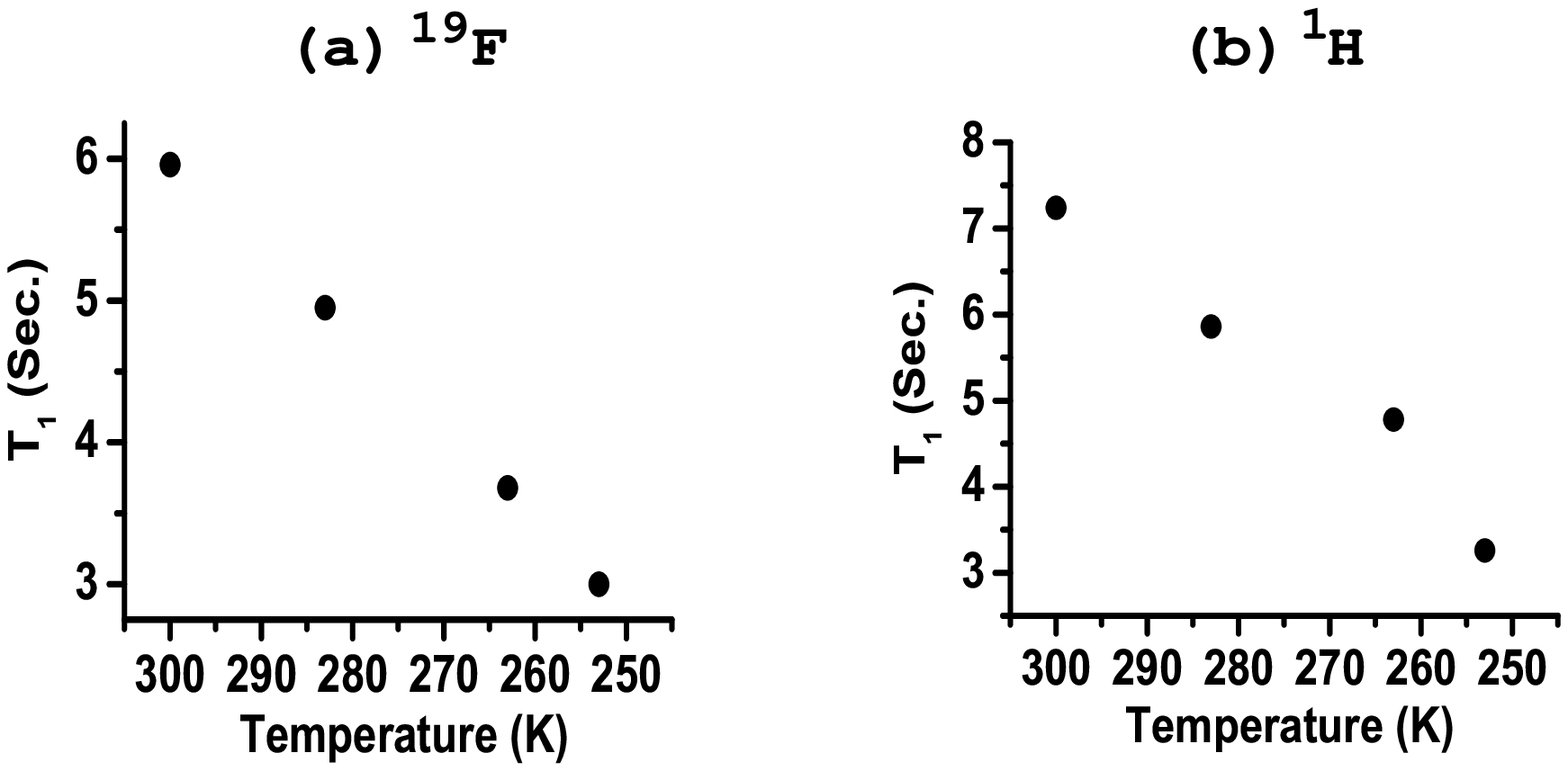,angle=0,width=19cm}
\caption{}
\label{FHT1}
\end{figure}
\end{center}

 Sample temperature was
varied to change the correlation time and hence the cross-correlation rate $\delta$. Four different sample 
temperatures, 300K, 283K, 263K and 253K were used. Fig.\ref{allspec} shows the proton and fluorine spectra obtained using recovery
measurement at four different temperatures. The spectra correspond to the initial PPS state and that after an interval of
2.5 sec. Fig.\ref{FHT1} shows the longitudinal relaxation times ($T_1$) of fluorine and
proton as function of temperature obtained from initial part of inversion-recovery experiment. A steady decrease in $T_1$ with decreasing temperature 
indicates that the dynamics of the
sample molecule is in the short correlation time limit \cite{abragam}. In this limit auto as well as cross-correlations increase 
linearly with decreasing temperature.\\ 

\begin{center}
\begin{figure}
\vspace*{-1cm}
\epsfig{file=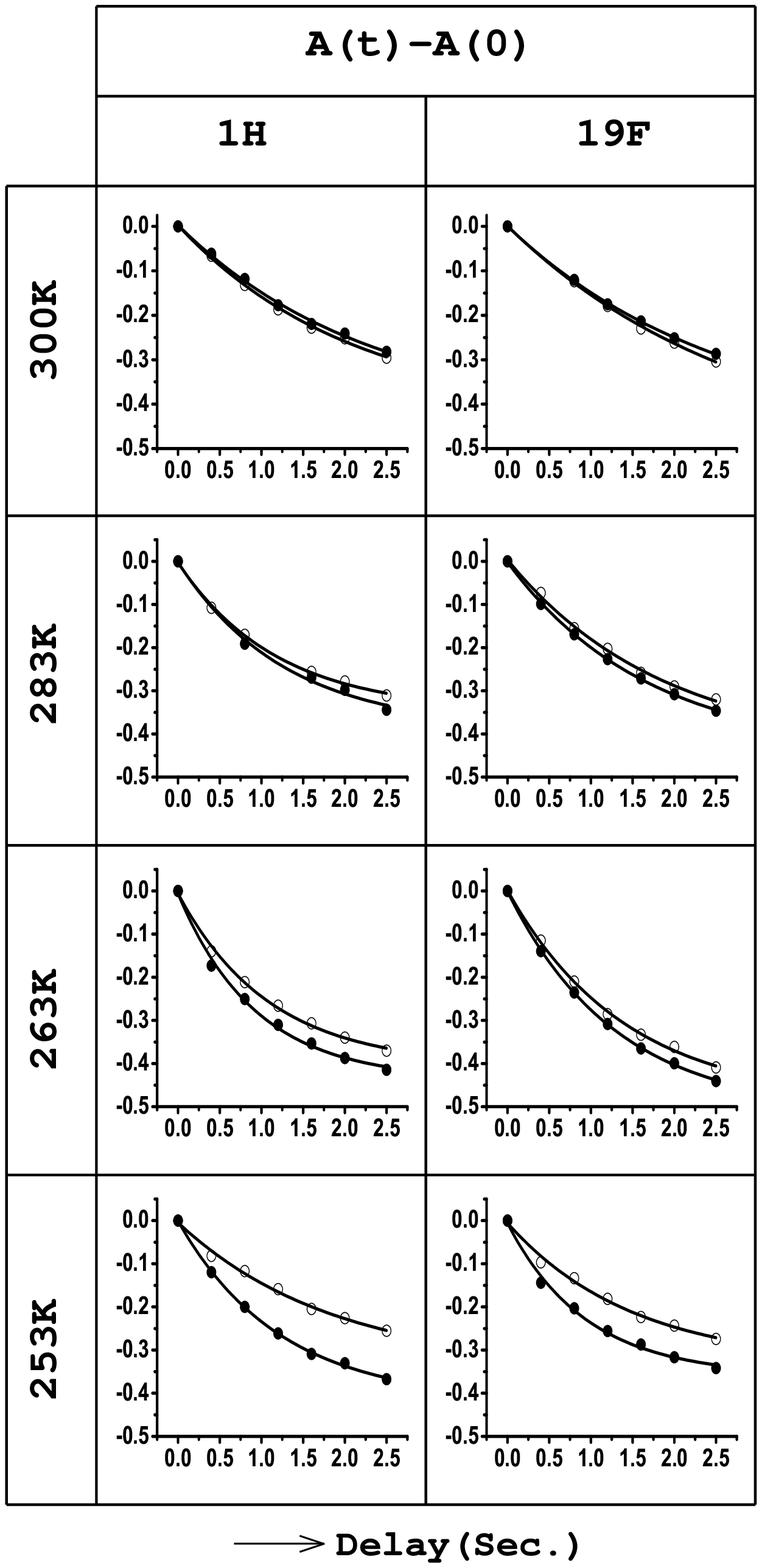,angle=0,width=11cm}
\caption{}
\label{Aplot}
\end{figure}
\end{center}

\begin{center}
\begin{figure}
\vspace*{-1cm}
\epsfig{file=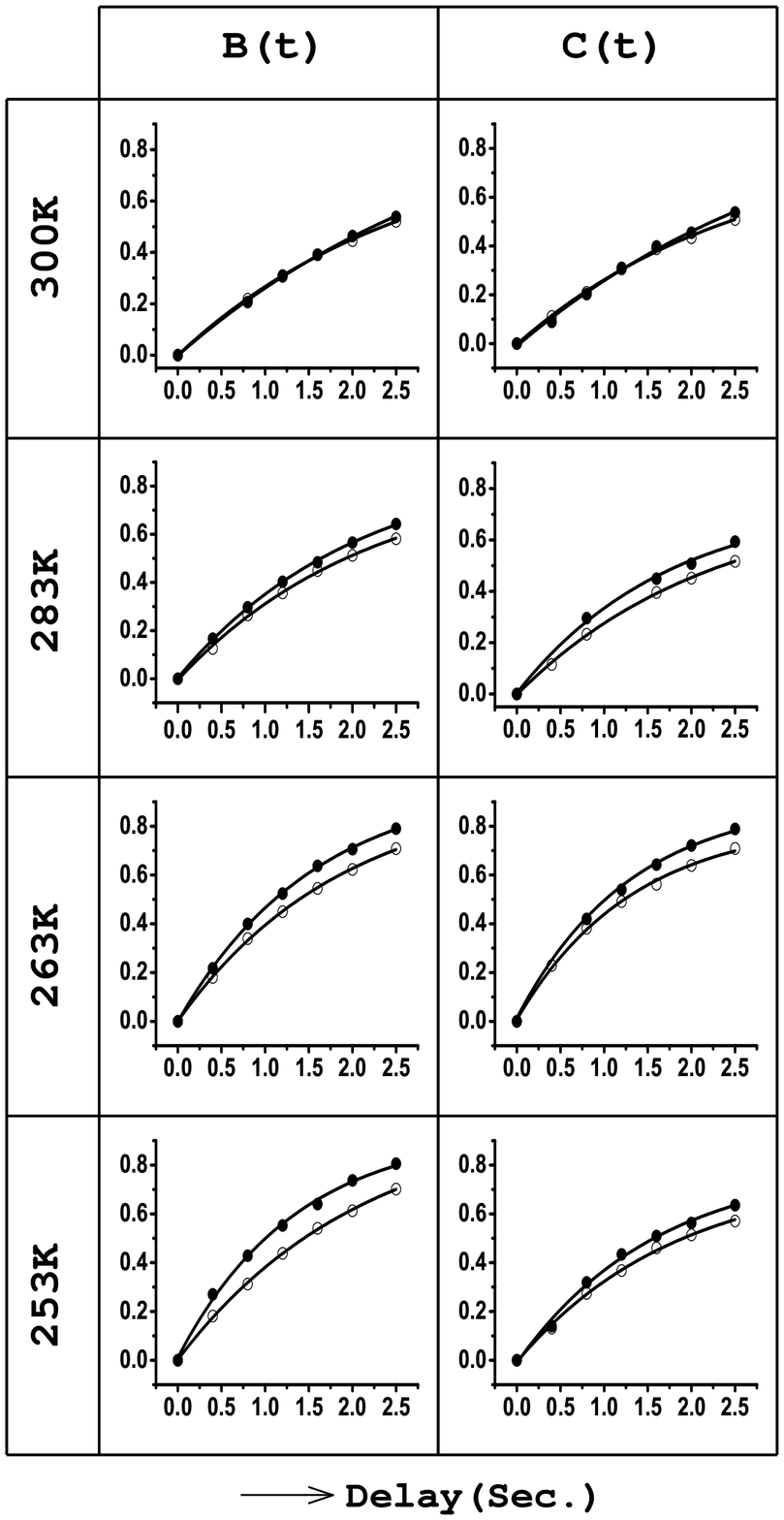,angle=0,width=11.5cm}
\caption{}
\label{BCplot}
\end{figure}
\end{center}

 All the spectra were fitted to bi-Lorentzian lines in MATLAB and various parameters were extracted using the Origin
software. Fig.\ref{Aplot} shows the decay of the coefficient $\mathcal{A}$ calculated independently from proton and fluorine spectra. 
At 300K, 
${\mathcal{A}}^{00}$ 
and ${\mathcal{A}}^{11}$ showed almost same rate of decay. As the temperature was gradually lowered, a steady increase in 
difference in decay rate was observed. This is due to the steady increase in cross-correlation rates with decreasing temperature, 
which is expected in the short correlation time limit.    
In Fig.\ref{BCplot} the growths of the coefficients $\mathcal{B}$ and $\mathcal{C}$ 
are shown. 
Similar to the coefficient $\mathcal{A}$, coefficients $\mathcal{B}$ and $\mathcal{C}$ also show differences in decay rate 
between $|00\rangle$ and $|11\rangle$ PPS at lower 
temperatures. The difference between ${\mathcal{B}}^{00}$ and ${\mathcal{B}}^{11}$ at any temperature was found to be larger 
compared to
between ${\mathcal{C}}^{00}$ and ${\mathcal{C}}^{11}$. This is expected since, according to Eq.\ref{abc-cross} the dominant 
cross-correlation 
factor in ${\mathcal{B}}^{00}$ and ${\mathcal{B}}^{11}$ is $\delta_{1,12}$ which is the cross-correlation between
CSA of fluorine with fluorine-proton dipolar interaction whereas in ${\mathcal{C}}^{00}$ and ${\mathcal{C}}^{11}$ the 
dominant factor is $\delta_{2,12}$ which is
cross-correlation between CSA of proton, which is much less than fluorine, with fluorine-proton dipolar interaction. 
Thus it is found that at lower temperatures the $|00\rangle$ PPS decays slower than the $|11\rangle$ PPS. The dominant 
difference in the decay
rates arises from the cross-correlations between the CSA of the fluorine and the dipolar interaction between the fluorine
and the proton spin. To the best of our knowledge this is the first study of its kind where the differential decay of the PPS has been
attributed to cross-correlations.     
\section{V. Conclusion}
We have demonstrated here that in samples having long $T_1$ cross-correlations plays an important role in determining the
rate of relaxation of pseudo pure state. In QIP sometimes one or more qubits having comparatively longer longitudinal
relaxation are used as storage or memory qubits. Recently Levitt et al. have demonstrated a long living antisymmetric state
arrived by shifting the sample from high to very low magnetic field, suggesting that this long living state could be used as memory
qubit \cite{levitthigh,levittlow}. In such cases fidelity of computation depends on how much
the memory qubits have been deviated from the initialized state at the beginning of the computation till the time they are
actually used. Theoretically it is shown here that in presence of cross-correlations, all the four PPS relax with different initial rates.
For positive cross-correlations the $|00\rangle$ PPS relaxes significantly slower than $|11\rangle$ PPS. It is
therefore important to choose a proper initial pseudo pure state according to the sample.    
\section*{Acknowledgments}
We gratefully acknowledge Prof. K. V. Ramanathan for discussions and Mr. Rangeet Bhattacharyya for his help in data processing. The use of DRX-500 high resolution liquid state 
spectrometer of the Sophisticated Instrument Facility, Indian Institute of
Science, Bangalore, funded by Department of Science and Technology (DST), New Delhi, is gratefully acknowledged. AK acknowledges "DAE-BRNS" for
"Senior Scientist scheme", and DST for a research grant.

\newpage
\begin{center}
{\bf FIGURE CAPTIONS}
\end{center}
Figure 1. (a) Chemical structure of 5-fluro 1,3-dimethyl uracil. The fluorine and the proton spins (shown by circles) are
used as the two qubits $I_1$ and $I_2$ respectively. (b) The energy level diagram of a two qubit system identifying the four states 00,01,10 and 11. Under high temperature and high
field approximation \cite{abragam} the relative equilibrium deviation populations are indicated in the bracket for each level. Assuming
this to be a weakly coupled two spin system the deviation populations become proportional to the gyromagnetic ratios $\gamma_1$ and 
$\gamma_2$. $I_{j}^{k}$ refers to the transition of the $j^{th}$ spin when the other spin is in state 
$|k\rangle$. Thus $H^0$ means the proton transition when the
fluorine is in state $|0\rangle$.\\\\
Figure 2. Population distribution of different energy levels of a two spin system in different pseudo-pure states. K is a constant whose 
value depends on the protocol used for the preparation of PPS. (a),(b),(c) and (d) show respectively the $|00\rangle$,$|01\rangle$,$|10\rangle$ and 
$|11\rangle$ PPS.\\\\
Figure 3. Schematic representation of decay of the coefficient $\mathcal{A}$ and growth of the coefficients $\mathcal{B}$ and $\mathcal{C}$. The magnetization modes
are normalized to their respective equilibrium values. In each sub-figure the three bars correspond to the modes $I_{1Z}$,$I_{2Z}$ and
2$I_{1Z}I_{2Z}$ from left to right. The amount of any mode present at any time is directly proportional to the height of the corresponding
bar. The numbers provided in the rightmost column represent typical values of the modes. (a) Thermal equilibrium. At thermal equilibrium
only $I_{1Z}$ and $I_{2Z}$ exist. (b) $|00\rangle$ pseudo-pure state 
just after creation, where 
all the three modes are equal in magnitude. For $|00\rangle$ PPS all modes are of same sign but this is not the case for other PPS 
[Eq.3]. Coefficient $\mathcal{A}$ is the common equal amount of all the modes and it is maximum at t=0.
(c) The amount of magnetization modes (schematic) at time $\tau$, after preparation of the PPS at t=0. The two single spin order modes 
increase and the two spin order mode decreases
from their initial values. (d) The state of various modes at time $\tau$, (same as fig.c) redrawn with filled bar to indicate the residual
value of $\mathcal{A}$. All the three coefficients $\mathcal{A}$,$\mathcal{B}$ and $\mathcal{C}$ are shown. 
$\mathcal{A}$ (shown by the filled bar), which is the measure of the PPS, has come down
by the same amount as the two spin order. $\mathcal{B}$ (shown by the empty bar) and $\mathcal{C}$ (shown by the striped bar) are the 
residual part of the 
single spin order modes $I_{1Z}$ and $I_{2Z}$
respectively.(e) The values of various modes and coefficients after a delay $\tau'>\tau$.     \\\\ 
Figure 4. Simulation of decay of coefficient $\mathcal{A}$. The boxes ($\boxempty$) and circles ($\circ$)
correspond to the $|11\rangle$ and $|00\rangle$ PPS respectively. In each plot deviation from initial value
($\mathcal{A}(t)-\mathcal{A}(0)$)has been
plotted.\\\\
Figure 5. Simulation of growth of coefficient $\mathcal{B}$ and $\mathcal{C}$. The boxes ($\boxempty$) and circles ($\circ$)
correspond to the $|11\rangle$ and $|00\rangle$ PPS respectively.\\\\

Figure 6. Relaxation of Pseudo pure state as monitored on (a) fluorine spin and (b) proton spin of the 5-fluro 1,3-dimethyl uracil at four
different temperatures. The top row in (a) and (b) show the equilibrium spectrum at each temperature. With decrease in temperatures the
lines broaden due to decreased $T_2$. The second row in (a) and (b) show the spectra corresponding to the $|00\rangle$ PPS, prepared by spatial
averaging method using J-evolution. The state of PPS was measured by $90^{o}$ pulse at each spin. The third row in (a) and (b) show the
spectra after an interval of 2.5 seconds after creation of the $|00\rangle$ PPS. The fourth row shows the spectra immediately after
creation of $|11\rangle$ PPS and the fifth row, the spectra after 2.5 seconds.\\\\\\   
Figure 7. Longitudinal relaxation time $T_1$ of fluorine (a) and proton (b) as function of temperature, measured from the
initial part of
inversion recovery experiment for each spin.\\\\
Figure 8. The deviation from initial value (at t=0) of the coefficient $\mathcal{A}$ of the PPS term calculated from Proton 
(left column) and Fluorine (right column) at four different sample temperature. The empty ($\circ$) and filled ($\bullet$) 
circles correspond to the $|00\rangle$ and $|11\rangle$ PPS respectively.\\\\
Figure 9. The growth of the coefficients $\mathcal{B}$ and $\mathcal{C}$ at different sample temperatures.
$\mathcal{B}$ was calculated from
Fluorine spectrum while $\mathcal{C}$ was calculated from the Proton spectrum. The empty ($\circ$) and filled ($\bullet$)
circles correspond to the $|00\rangle$ and $|11\rangle$ PPS respectively.\\\\
\newpage

\newpage

\end{document}